\documentclass[12pt]{iopart}
\usepackage{CJK}
\usepackage[T1]{fontenc}
\usepackage[latin9]{inputenc}
\usepackage{times}
\usepackage[sort&compress,numbers]{natbib}
\usepackage{color} 
\usepackage{xspace}
\usepackage{amsbsy}
\usepackage[pdftex]{graphicx}
\usepackage{bm}
\usepackage{float}
\usepackage{pict2e}
\usepackage{amssymb}
\usepackage{adjustbox}
\usepackage{epstopdf}

\usepackage[unicode,breaklinks]{hyperref}
\hypersetup{ 
    unicode=true,
    plainpages=false, 
    colorlinks=true,
    linkcolor=blue,
    citecolor=blue,
    filecolor=black,
    urlcolor=blue
}
\usepackage{color}
\usepackage{url}
\usepackage{verbatim}

\begin{document}

\begin{CJK*}{GBK}{ }
\title[Percolation in a spinor condensate]{Domain percolation in a quenched ferromagnetic spinor condensate} 
\author{Nanako Shitara$^{1,2}$, Shreya Bir$^{1,3}$, and P.~Blair Blakie$^{1,4}$}  
\address{1. Dodd-Walls Centre for Photonic and Quantum Technologies, Department of Physics, University of Otago,  PO Box 56, Dunedin 9016, New Zealand}
\address{2. Department of Physics, University of California, Berkeley, CA 94720, USA}
\address{3.  Otago School of Medical Sciences, University of Otago, Dunedin, New Zealand}

\address{4. Beijing Computational Science Research Center, Beijing 100094, China}
\ead{blair.blakie@otago.ac.nz}
\end{CJK*}
\begin{abstract}  
  {We show that the early time dynamics of easy-axis (EA) magnetic domain formation in a spinor condensate is described by percolation theory. These dynamics could be initialized using a quench of the spin-dependent interaction parameter. We propose a scheme to observe the same dynamics by quenching the quadratic Zeeman energy and applying a generalized spin rotation to a ferromagnetic spin-1 condensate.}  Using simulations we investigate the finite-size scaling behaviour to extract the correlation length critical exponent and the transition point. We analyse the sensitivity of our results to the early-time dynamics of the system, the quadratic Zeeman energy, and the threshold condition used to define the positive (percolating) domains. 
\end{abstract}
 
\maketitle

\pagestyle{plain}
\tableofcontents

\section{Introduction}
Spinor condensates have spin and gauge degrees of freedom and are novel systems for exploring various types of symmetry breaking phase transitions in an isolated and highly controllable quantum system \cite{Kawaguchi2012R,StamperKurn2013a}. 
For more than a decade a range of beautiful experiments have explored phase transitions and associated dynamics in spinor condensates \cite{Sadler2006a,Leslie2009a,Liu2009a,Vengalattore2010a,Guzman2011a,Bookjans2011b,Jacob2012a,Vinit2013a,Jiang2014a,Kang2017a}. 
Much theoretical work has been concerned with defect formation arising from how the phase transition is crossed (e.g.~see \cite{Saito2007a,Uhlmann2007a,Lamacraft2007a,Sau2009a,Barnett2011,Witkowska2013a,Witkowska2014a}), and on the universal growth laws 
 describing how the domains anneal long after the system has passed through the phase transition (e.g.~see \cite{Mukerjee2007a,Kudo2013a,Kudo2015a,Williamson2016a,Symes2017b}).

  {In this work we focus on the geometrical-statistical properties of the spin domains that form in a spinor condensate prepared in an unstable initial state using a generalized quench. These spin domains emerge in the easy-axis (EA) phase of a ferromagnetic spin-1 condensate, and prefer to have their magnetization either aligned (positive) or anti-aligned (negative) with the external magnetic field. The domains initially grow randomly, seeded by quantum or thermal noise.} We analyse these domains in terms of percolation theory, canonically formulated to describe the behaviour of connected clusters in a random graph. 
It is useful to briefly introduce the central idea of percolation theory (e.g.~see \cite{Stauffer1971}). The so called site percolation problem involves a lattice in which sites are randomly and independently occupied with a probability $p$. The central question of percolation theory is for a given $p$, what is the probability that a path (of occupied sites) exists extending across the lattice? Such a path is also called a percolating cluster. For an infinite lattice there is a critical value $p_c$ (the percolation threshold) such that for $p<p_c$ a percolating cluster never occurs, while for $p>p_c$ it always occurs. It is found that a ``geometrical'' phase transition occurs at $p_c$, such that the geometric properties of the clusters near $p_c$ are characterized by universal critical exponents \cite{Bunde1991}. 

The mapping of the continuous domains that occur in the EA phase of a spinor condensate onto a site percolation problem is one of the issues we discuss in this paper. Our basic approach is to consider positively magnetized regions as ``occupied'' sites, and to then analyse whether there exists a contiguous positive domain that spans the system. 
In order to explore the percolation behaviour it is necessary to vary the  effective $p$ value, i.e.~relative proportion of the system occupied by positive domains. We demonstrate that this can be done using a generalized spin rotation to modify the initial state magnetization. The particular rotations we introduce are also chosen to reduce heating that will occur as the EA domains form. 

  Takeuchi \textit{et al.}~showed that the domains forming  in the immiscibility transition of a binary condensate are described by percolation theory \cite{Takeuchi2015a}.  The EA ferromagnetic phase of a spinor condensate and immiscible phase of a binary condensate also have been revealed to have similar behaviour in phase ordering dynamics (c.f.~\cite{Hofmann2014,Williamson2016a,Takeuchi2016a,Andreane2017a,Takeuchi2017a}). Spinor systems have some potential advantages for exploring phase transition dynamics, such as ease of preparing initial states and techniques for directly probing the spin degrees of freedom.

We now briefly outline the paper. In Sec.~\ref{SecFormalism} we introduce the basic formalism for describing the dynamics of a spin-1 condensate. We present the initialization procedure we have developed for the quench dynamics to produce systems in which the relative proportion of positive domains can be controllably varied. We also introduce the numerical scheme we use to simulate the dynamics of a uniform quasi-two-dimensional (quasi-2D) system.  Section \ref{Sec:Results} contains the main results of the paper. We define an effective occupation probability $p$ in terms of the conserved system magnetization, and measure percolation by identifying domains that span or wrap around the periodic simulation grid. We study the percolation behaviour of the system using an ensemble of simulations to evaluate the probability that percolating domains occur. We perform a finite-size scaling analysis to demonstrate how the percolation probabilities change as the system size increases. This allows us to accurately extract the correlation length critical exponent $\nu$, and extrapolate to the infinite system percolation threshold $p_c$. We show that this threshold is independent of the percolation measure (i.e.~spanning or wrapping domains), but is sensitive to the time after the quench, the quadratic Zeeman energy, and the condition used to identify the positive domains.  In Sec.~\ref{SecConclusion} we conclude and discuss the outlook for this work. Since our main results of the paper are presented for a uniform quasi-2D system we also briefly touch on the feasibility for observation in experiments and present a result for experimentally realistic parameters.

\section{Formalism and quench}\label{SecFormalism}
\subsection{Hamiltonian and evolution} 
We study a homogeneous quasi-2D ferromagnetic spin-1 condensate. Such a system can be realized with $^{87}$Rb atoms prepared in the ground state $F=1$ hyperfine manifold and confined in an optical trapping potential.
The system is then described by the spinor field $\bm{\psi}=(\psi_1,\psi_0,\psi_{-1})$, where $\psi_m$ is the component of the system in the $m$  Zeeman sublevel.

The dynamics of this system is described by the spin-1 Gross-Pitaevskii equation (GPE)
\begin{eqnarray}\label{spinGPEs}
i\hbar\frac{\partial\bm{\psi}}{\partial t}=\left(-\frac{\hbar^2\nabla^2}{2M}+qf_z^2+g_nn+g_s\bm{F}\cdot\bm{f}\right)\bm{\psi},
\end{eqnarray}
where $q$ is the quadratic Zeeman energy shift, $g_n$ is the positive coupling constant for the density dependent interaction, and  $g_s$ is the negative  (i.e.~ferromagnetic)  coupling constant for the spin dependent interaction \cite{Ho1998a,Ohmi1998a}. We have also introduced
\begin{eqnarray}
n(\mathbf{x})&=\sum_{m}|\psi_m(\mathbf{x})|^2, \\
F_\nu(\mathbf{x})&=\sum_{m,m'}\psi_{m'}^*(\mathbf{x})(f_\nu)_{m'm}\psi_m(\mathbf{x}),\qquad \nu=\{x,y,z\},
\end{eqnarray}
as the total (areal) density and the $\nu$-component of spin density, where $\mathbf{f}\equiv(f_x,f_y,f_z)$ are the spin-1 matrices, and $\mathbf{F}=(F_x,F_y,F_z)$ is the spin density vector. The effect of the linear Zeeman shift is trivially removed by moving to a frame rotating at the Larmor frequency, and is neglected in Eq.~(\ref{spinGPEs}).  
Evolution according to the GPE (\ref{spinGPEs}) conserves energy and total particle number. As the system is axially symmetric in spin space (invariant to spin rotations about $z$), the $z$-magnetization \begin{equation}
M_z\equiv\int d^2\mathbf{x}\,F_z,
\end{equation}
is also a constant of motion. The magnetization controls the effective proportion of positive and negative domains in the EA phase, and in the next subsection we consider how to alter $M_z$ in a way suitable for exploring the percolation transition.

\subsection{Initial state preparation and quench}\label{SecInitState}

  { 
The spin domain formation dynamics we consider here can be implemented as a quench in which a single parameter of the system is changed and the system transitions towards a new equilibrium state. 
The initial state we consider is the uniform (zero-momentum) spinor
\begin{eqnarray}
\bm{\psi}_0=\sqrt{\frac{n_0}{2}}\left(\begin{array}{c} \cos \varphi +\sin \varphi \cr  0 \cr \sin \varphi -\cos \varphi \end{array}\right),\label{EqInitStateQ2}
\end{eqnarray}
where $n_0$ is the condensate density, and $\varphi$ is an angle parameterizing the state. This miscible two-component state is the ground state for a spin-1 condensate with anti-ferromagnetic interactions ($g_s>0$) at $q<0$. The quench of interest is to suddenly change $g_s$ to a negative value, whereby the two components are immiscible and EA domains will form.
 In principle the sign of $g_s$ can be changed using optical \cite{Fatemi2000a} or microwave-induced Feshbach resonances \cite{Papoular2010a}. However, as these techniques have not yet been demonstrated in spinor experiments, we propose a different scheme to equivalently initialize the dynamics in a ferromagnetic condensate. While (\ref{EqInitStateQ2}) is not an equilibrium state for a ferromagnetic condensate, } it can be prepared starting from a polar state $\bm{\psi}_P\sim[0,1,0]^T$, which is the unmagnetized ground state for large positive $q$, by driving the atomic internal states using two subsequent electromagnetic pulses. The first pulse is the spin-1 rotation $e^{-i\frac{\pi}{2}f_y}$, which produces the intermediate state $\bm{\psi}'\sim[-1/\sqrt{2},0,1/\sqrt{2}]^T$ (e.g.~see \cite{Seo2015a,Seo2016a}). The second pulse is the pseudo-spin half rotation $e^{-i\varphi\sigma_y}$ of variable angle $\varphi$, performed on the $m=\pm1$ levels, where $\sigma_y$ is the $y$-Pauli spin matrix. The result of both pulses is then the desired initial state (\ref{EqInitStateQ2}). The second rotation could be driven as a two-photon transition between the $m=\pm1$ states using microwave radiation detuned from the intermediate $|F=2, m=0\rangle$ state.
We emphasize that the atomic physics toolbox of coherent manipulations presents various ways to quickly and reliably engineer (\ref{EqInitStateQ2}), also see \cite{Smith2013a}.

We briefly comment on the motivation for initializing the system to this particular initial state. First, this state has a magnetization  of $M_z = N\sin2\varphi$ controlled by the angle of the spin rotation, where $N=\int d^2\mathbf{x}\,n$ is the total number of particles. Second, particularly in the regime of small rotations ($|\varphi|\ll \pi/2$) of interest here, this state is close to the state $[-1/\sqrt{2},0,1/\sqrt{2}]^T$, which undergoes less heating\footnote{Atoms   $m=0$  liberate energy of $-q$ when they spin mix into the $m=\pm1$ levels to form EA domains.} during the formation of EA domains (see discussion in Ref.~\cite{Andreane2017a}), and thus produces cleaner domains.

\subsection{Simulations}
It is useful to introduce   $q_0=2|g_s|n_0$ as characteristic spin energy, and use it to define the spin time\footnote{  {This is the characteristic timescale over which spin can evolve when driven by the spin-dependent interaction.}} $t_s\equiv \hbar/q_0$, and the spin healing length $\xi_s\equiv\hbar/\sqrt{q_0 M}$. 
 Simulations are performed using the spin-1 GPE with weak noise (representing the vacuum fluctuations) added to seed the dynamic instabilities. The noise is added to the polar condensate state as described in Ref.~\cite{Williamson2016b} and then the spin rotations described in the previous subsection are applied to this state to prepare the initial condition. The simulation is performed on a 2D square grid of spatial dimension $L\times L$ covered by a grid of $N_L\times N_L$ equally spaced points, with periodic boundary conditions. 
The GPE is evolved using the fourth order symplectic technique described in Ref.~\cite{Symes2016a}.

\section{Results}\label{Sec:Results}
\subsection{Percolating species and effective $p$ value}
The initial condition (\ref{EqInitStateQ2}) is unstable for a condensate with ferromagnetic interactions. For $q<0$ the ground state of the system is an EA ferromagnetic state, i.e.~a phase in which the ground state condensate maximally aligns or anti-aligns along the spin $z$ axis  (i.e.~the system prefers to have $F_z\approx\pm n_0$). For a spatially extended system the initial condition will evolve into the EA phase by producing small domains, which can be labelled as positive and negative by the sign of $F_z$. Our interest here is to characterise the properties of these domains.
To map the EA system onto a percolation problem we will consider positive domains to be ``occupied''. 
In practice due to domain walls, and heating from the quench, the system is not perfectly polarized. We take the positive domains to be specified by the set of points
\begin{eqnarray}
\sigma_+(\epsilon)=\{\mathbf{x}: F_z(\mathbf{x})>\epsilon n_0\},\label{eq:sigmadefn}
\end{eqnarray}
where  $\epsilon\ge0$ is a constant that defines the threshold condition for a point to be included in a positive domain. We denote the total area of $\sigma_+$ as $A_+$, i.e.,~the total area of positive domains. 
The ratio $A_+/L^2$ quantifies the probability that a randomly selected point in the system is in a positive domain, and would be a useful choice for the occupation probability $p$ used in standard percolation theory. However, $A_+$ is not a constant of motion and changes during evolution.  We instead choose to use
\[
p=\frac{1}{2}\left(\frac{M_z}{N}+1\right),\label{Eqpval}
\]
as an effective $p$ value, defined in terms of the constant of motion $M_z$. The value of $M_z$ and hence $p$ is set by the initial state (\ref{EqInitStateQ2}) using the spin rotation angle $\varphi$.
For sharp domain walls and perfectly polarized condensate domains (i.e.~everywhere $F_z=\pm n_0$), then this definition is equivalent to $A_+/L^2$. 
 \begin{figure}[htbp] 
   \centering
    \includegraphics[width=6in]{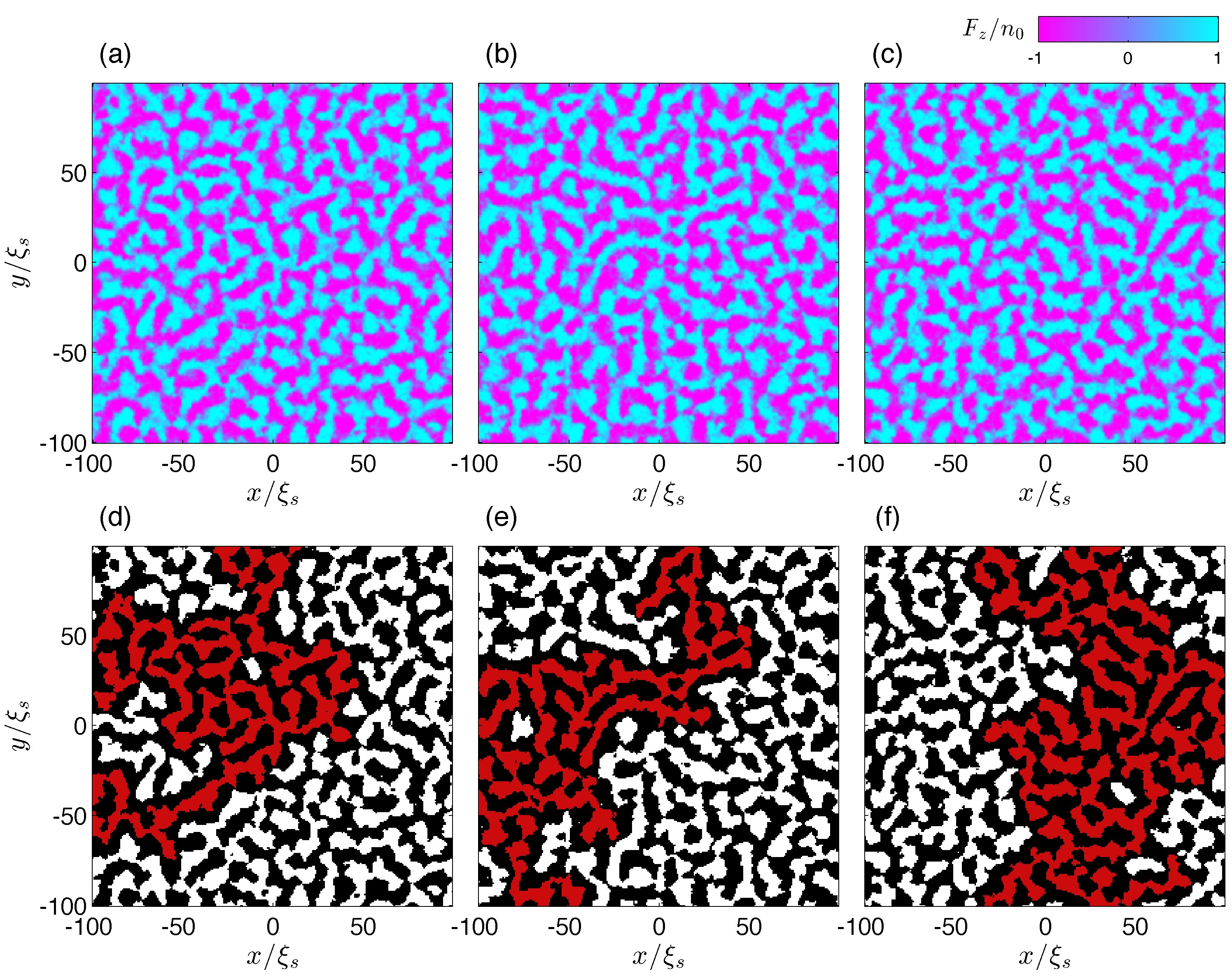} 
     \caption{Examples of domains formed after a quench into EA phase. 
     (a)-(c)  Three realizations of the $z$-spin density at $t=50\,t_s$, which differ only in the seed noise added to the initial condition. 
     (d)-(f) The respective binary images corresponding to (a)-(c) [see text]. Occupied sites (positive domains) are shaded white, except for the largest domain which is shaded red.  
     Simulations for $n_0=10^4/\xi_s^2$, $q=-0.3\,q_0$, $g_s=-\frac{1}{3}g_n$, $L=200\,\xi_s$, $N_L=256$, $p=0.5$ and analyzed with a density threshold parameter of $\epsilon=0.1$.}
   \label{fig:Domains}
\end{figure}

\subsection{Application of percolation analysis}\label{Secpercanalysis}
In Figs.~\ref{fig:Domains}(a)-(c) we present examples of the $F_z$ density at a time of $t= 50\,t_s$ after a quench with $p=0.5$ (i.e.~$M_z=0$) for three different realizations for initial noise. In the simulations the initial magnetization is uniform ($F_z=0$ for $p=0.5$ in Fig.~\ref{fig:Domains}), but dynamic instabilities lead to the formation of EA domains. These domains become sufficiently well formed by   around $t\sim10\,t_s$ that after this time we can analyse their properties. 
 To perform this analysis we apply the threshold condition (\ref{eq:sigmadefn}) at every computational grid point to construct a binary image: points satisfying the threshold condition are assigned a value of 1 and all other points are assigned a value of 0. This allows us to calculate the total domain area as $A_+= \mathcal{N}_+(\Delta x)^2$, where $\mathcal{N}_+$ is the total number of  points with a value of 1, and $\Delta x=L/N_L$ is the grid point spacing. We then apply the Hoshen-Kopelman algorithm \cite{Hoshen1976a} to the binary image to enumerate distinct positive domains (i.e.~sets of occupied points in connected clusters). Our main analysis is to perform a quench simulation and then  
at some final time $t$ quantify if any domain percolates in the following ways:
\vspace*{-0.4cm}\paragraph{Spanning domain:} A domain which touches two opposite sides of the system (i.e.~left to right, or top to bottom).
\vspace*{-0.4cm}\paragraph{Wrapping domain:} A spanning domain whose spanning ends overlap when periodic boundary conditions are imposed.

  {We emphasize that the definition of a percolating domain is not uniquely defined, and other definitions could be used. Our results will show that while these two choices are different for a finite system, they predict the same percolation threshold for an infinite system. }

\vspace*{0.3cm}
Examples of binary images are shown in Figs.~\ref{fig:Domains}(d)-(f), with the largest domains indicated. These three examples demonstrate cases where the largest domain:  (d) neither spans nor wraps; (e) spans, but does not wrap; (f) spans and wraps.   By running a large number ($N_{\mathrm{traj}}$) of independent trajectories for a given $p$ value, each with different seed noise, we can evaluate the probability that a domain will span or wrap under those conditions as 
\begin{equation}
P_{\mathrm{span}}=\frac{N_{\mathrm{span}}}{N_{\mathrm{traj}}},\qquad P_{\mathrm{wrap}}=\frac{N_{\mathrm{wrap}}}{N_{\mathrm{traj}}},
\end{equation}
where $N_{\mathrm{span}}$ ($N_{\mathrm{wrap}}$) is the number of trajectories where at least one spanning (wrapping) domain was found. Since all wrapping domains are spanning domains, we have that $P_{\mathrm{wrap}}\le P_{\mathrm{span}}$.

  \begin{figure}[htbp] 
   \centering
   \includegraphics[width=4.0in]{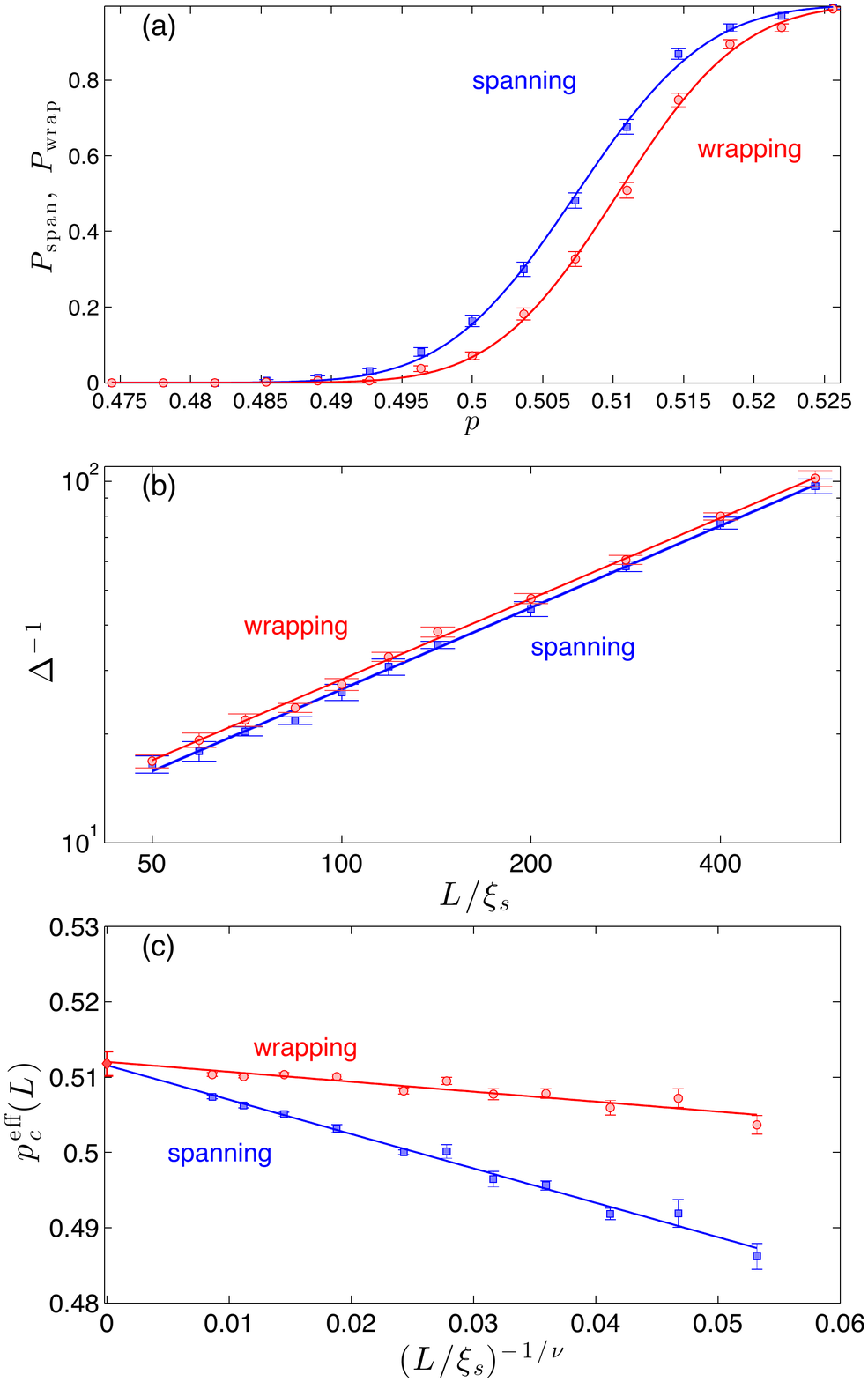} 
   \caption{Analysis of domain percolation at $t=50\,t_s$.
   (a) Probability that at least one domain is spanning or wrapping for a system of size $L=200\,\xi_s$ with $N_L=256$ points in each direction. 
   Lines are the fits to the numerical results using Eq.~(\ref{Eq:pfit}). 
   (b) Finite size scaling of the inverse percolation transition width with system size revealing  $1/\nu$. Lines of best fit give  $\nu=1.34\pm0.04$.
   (c) Finite size scaling of the effective percolation threshold $p^{\mathrm{eff}}_c(L)$.  Both sets of results extrapolate to the same infinite system threshold value $p_c=0.512\pm0.002$ (red diamond).
   Results are computed using  $N_{\mathrm{traj}}=590$ for $q=-0.3\,q_0$, $g_s=-\frac{1}{3}g_n$ and $n_0=10^4/\xi_s^2$. The grid spacing is held fixed at $\Delta x=0.78\,\xi_s$ as $L$ and $N_L$ are changed.  
   }
   \label{fig:t50}
\end{figure}

In Fig.~\ref{fig:t50}(a) we plot the spanning and wrapping probabilities as a function of $p$, calculated using $N_{\mathrm{traj}}=590$ simulation trajectories for each $p$ value.
These probabilities are monotonically increasing functions of $p$ and have a sigmoidal shape. This indicates a percolation transition, rounded off by the finite size (here $L=200\,\xi_s$) of the simulation. To analyse these results we fit the probabilities to the function 
\begin{equation}
P_{\mathrm{fit}}(p)=\frac{1}{2}\left[1+\mathrm{erf}\left(\frac{p-p^{\mathrm{eff}}_c(L)}{\Delta(L)}\right)\right],\label{Eq:pfit}
\end{equation}
[fits shown in Fig.~\ref{fig:t50}(a)] to determine the  {effective percolation threshold} $p^{\mathrm{eff}}_c(L)$ and the  {width of the percolation transition} $\Delta(L)$. 

We can perform similar simulations and analysis to that in Fig.~\ref{fig:t50}(a) but for simulations of different systems sizes $L$. From the fits to these results we can then determine the finite size scaling of $p^{\mathrm{eff}}_c(L)$ and $\Delta(L)$.
The scaling result for the transition width 
 \begin{equation}
 \Delta(L)\propto L^{-1/\nu},\label{EqDeltaFS}
 \end{equation}
allows us to extract the $\nu$ critical exponent. In standard percolation theory this critical exponent describes the divergence of the correlation length $\xi$ as the percolation transition is approached: $\xi\sim|p-p_c|^{-\nu}$. The scaling relationship (\ref{EqDeltaFS}) usually furnishes a more accurate value for $\nu$ than would be obtained by measuring the correlation length divergence \cite{Rintoul1997a}.
In Fig.~\ref{fig:t50}(b) we plot $\Delta(L)^{-1}$ versus $L$ for systems of 11 different sizes from $L=50\xi_s$ up to $565.6\xi_s$ (at fixed grid point spacing). The results for $\Delta(L)^{-1}$ demonstrate that for both spanning and wrapping percolation measures the transition sharpens (i.e.~$\Delta$ narrows) as the system size increases. The best power-law fits to both sets of results are shown as straight lines in Fig.~\ref{fig:t50}(b), and the inverse slopes of both lines are the same within error bars giving  $\nu=1.34\pm0.04$ [c.f.~Eq.~(\ref{EqDeltaFS})]. This result is in agreement with the value of $4/3$ expected for standard percolation in 2D.

 We can now examine the percolation threshold $p_c$, i.e.~the $p$ value in the infinite system where the probability for finding a percolation domain abruptly changes from zero to unity.   
Finite size scaling predicts that the difference between the finite system percolation threshold and the infinite threshold $p_c$ scales as 
\begin{equation}
p^{\mathrm{eff}}_c(L)-p_c \propto L^{-1/\nu}. 
\end{equation}
Using the value of $\nu$ extracted above, we  plot $p^{\mathrm{eff}}_c(L)$ versus $L^{-1/\nu}$ in Fig.~\ref{fig:t50}(c). The results for both percolation measures are well fitted by straight lines. These lines are of different slopes, but extrapolate to the same $y$-intercept corresponding to the infinite system limit (i.e.~$L^{-1/\nu}\to0$) of $p_c=0.512\pm0.002$. This shows that while in a finite size system the various percolation measures are distinct, in the infinite system limit the percolation transition is sudden and insensitive to the particular definition we use for a percolating domain.

\begin{figure}[htbp] 
   \centering 
    \includegraphics[width=4.25in]{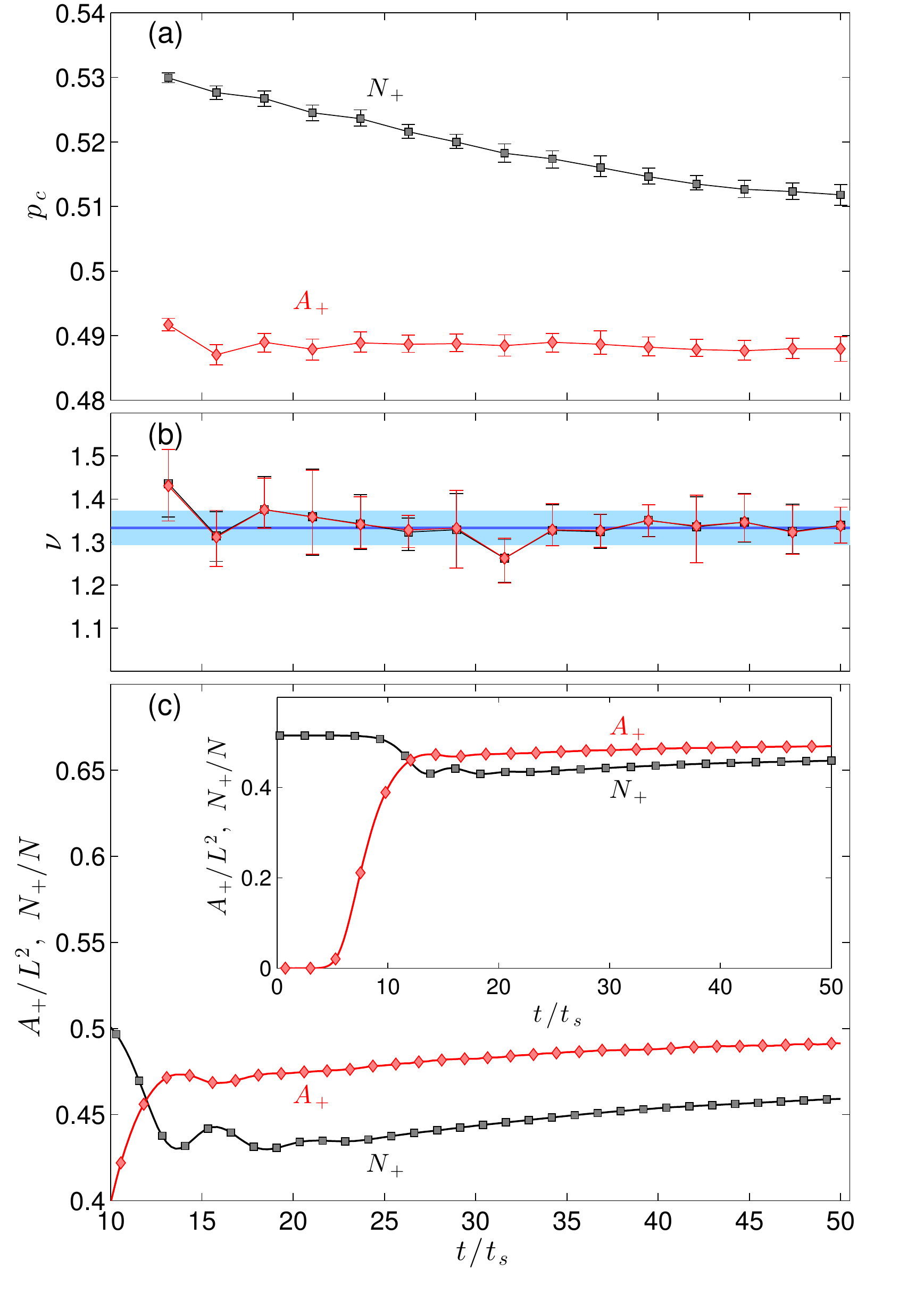} 
   \caption{Results of percolation analysis in the early-time evolution of the spinor condensate.
   (a) Extrapolated critical value  $p_c$ calculated using $p$ [Eq.~(\ref{Eqpval})]  (squares) and using the domain area $A_+$  (diamonds).
   (b) The value of the $\nu$ exponent extracted from results as in Sec.~\ref{Secpercanalysis} [same symbols as in (a)].  Horizontal line indicates the value $\frac{4}{3}$, and shaded region indicates a $\pm0.04$ error bar. (c) Evolution of the  relative population of $m=1$ atoms (i.e.~$N_+/N$) and the relative positive domain area (i.e.~$A_+/L^2$) for a simulation with $L=400\,\xi_s$ and $p=0.515$.  Inset: same results plotted with different range to show the initial ($t<10\,t_s$) behaviour. Results analyzed using the density threshold condition $\epsilon=0.1$.}
   \label{fig:td_domains}
\end{figure}

\subsection{Generality of results}
The analysis of the last subsection showed that the formation of percolating EA domains exhibits a well-defined transition in the infinite system, consistent with the standard 2D percolation transition. However, in order to better understand the generality of this result we investigate how the percolation behaviour changes with time [Sec.~\ref{Sectimeevolv}], its dependence on $q$ [Sec.~\ref{Secq}], and sensitivity to the density threshold condition $\epsilon$ [Sec.~\ref{Sece}].

\subsubsection{Early-time evolution}\label{Sectimeevolv}

We can repeat the analysis summarized in Fig.~\ref{fig:t50} for other times after the domains form (i.e.~for $t>10\,t_s$). The results are presented in Figs.~\ref{fig:td_domains}(a) and (b), showing that the exponent $\nu$ is generally insensitive to the time evolution, whereas the threshold value $p_c$ decreases with time. The dominant cause of this time dependence is spin mixing, i.e.~the process where atoms in  $m=1$ and $m=-1$ collide and both convert into $m=0$ atoms (or the time reversed process). We show the population of the $m=1$ level $N_+=\int d^2\mathbf{x}\,|\psi_1|^2$ in Fig.~\ref{fig:td_domains}(c), which is seen to dip down and briefly oscillate at around $t\approx15\,t_s$, and then slowly increase back towards its initial value as time progresses. Spin mixing is more significant at small negative $q$ values where the $m=0$ level is energetically accessible [we discuss the role of  $q$ further in Sec.~\ref{Secq}]. 
 
The tendency of the positive domain to percolate is related to the total domain area, and hence $N_+$. The reduction in $N_+$ occurring when the domains first form, means that a higher initial $N_+$ population is needed to see percolation, or equivalently a higher magnetization (i.e.~$p$ value). For this reason $p_c$ is higher early on and decreases with time as the $N_+$ population increases.

It is also useful at this point to return to the definition of the $p$ value given in Eq.~(\ref{Eqpval}).  An alternative definition for the occupation probability $p$ is the fraction of the system covered by positive domains, i.e.~
\begin{eqnarray}
p'(t)=\frac{A_+}{L^2}.
\end{eqnarray}
We have explicitly given this quantity a $t$ dependence to indicate that unlike Eq.~(\ref{Eqpval}) this quantity is not a constant of motion. An example of the evolution of $A_+$ is given in Fig.~\ref{fig:td_domains}(c), showing that it rapidly grows from zero as the domains initially form, and then more slowly at later times as $N_+$ increases. We can adapt the analysis of our percolation results by plotting the probabilities at each time $t$ against $\langle p'(t)\rangle$ [c.f.~Fig.~\ref{fig:t50}(a)], where $\langle p'(t)\rangle$ is the average over the $N_{\mathrm{traj}}$ trajectories of the value of $p'$ at time $t$. Performing fits and finite size scaling analysis (as described in Sec.~\ref{Secpercanalysis}) we can then extract a percolation threshold $p'_c$ and the critical exponent $\nu$. The results of this analysis, applied to the same trajectories analyzed above in terms of $p$ is also shown in Fig.~\ref{fig:td_domains}(a) and (b). We see that the percolation threshold $p_c'$ is now essentially constant at a value of $p'_c\approx0.49$. The critical exponent values are almost identical to the previous analysis.

\begin{figure}[htbp] 
   \centering 
    \includegraphics[width=4.75in]{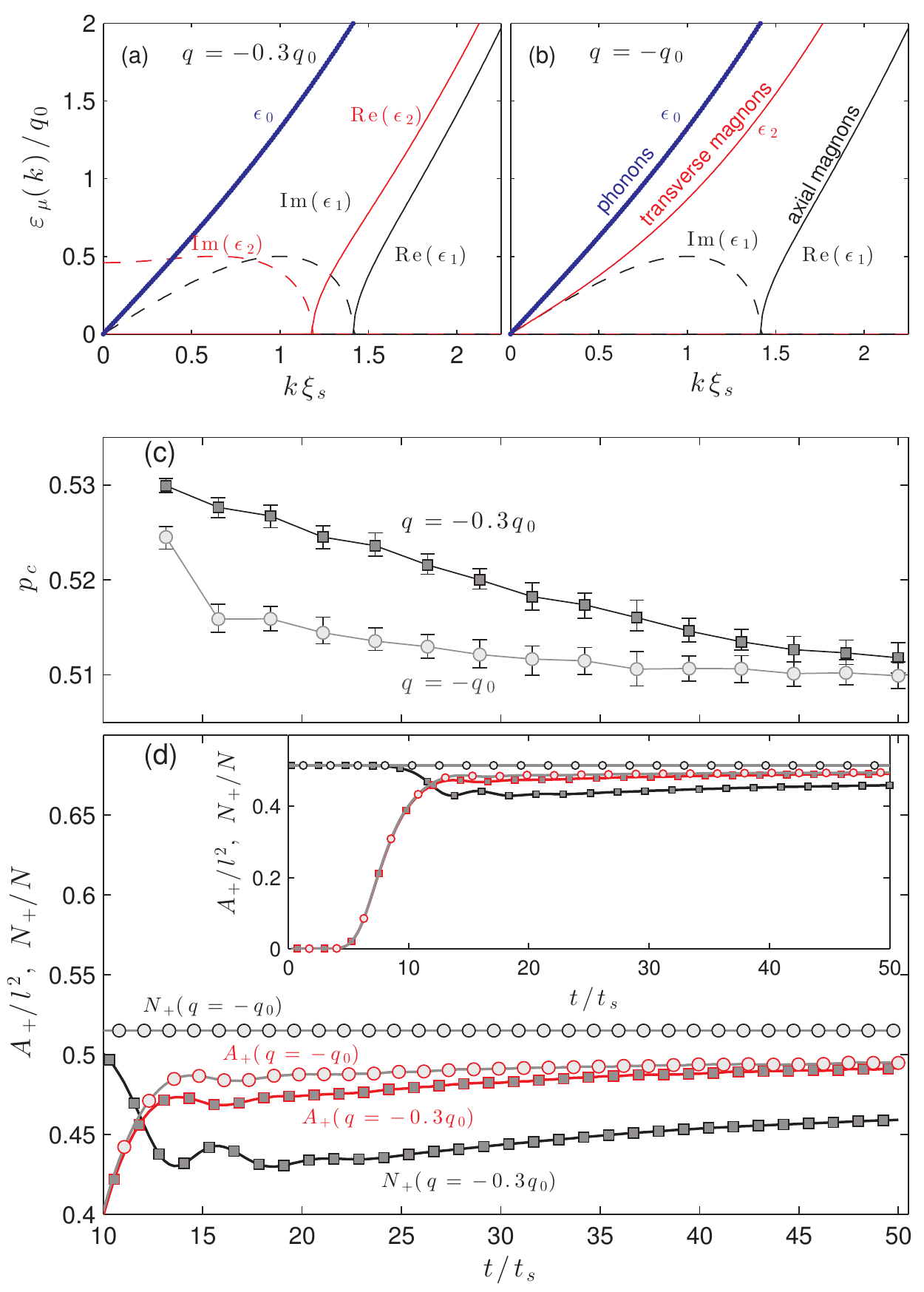} 
   \caption{Effect of changing $q$. The dispersion relation of excitations on the initial condensate (\ref{EqInitStateQ2}) with $p=0.5$ initial state for (a) $q=-0.3q_0$ and (b) $q=-q_0$. The dynamically unstable modes have imaginary parts (dashed lines) causing them to grow following the quench. The character of the modes is labelled in subplot (b).
   (c) The extrapolated critical value  $p_c$ calculated using $p$ [Eq.~(\ref{Eqpval})]  for $q=-0.3q_0$ (squares) [as in Fig.~\ref{fig:td_domains}(a)] and for $q=-q_0$ (circles).
   (d)   Evolution of the  relative population of $m=1$ atoms (i.e.~$N_+/N$) and the relative positive domain area (i.e.~$A_+/L^2$) for a simulation with $L=400\,\xi_s$ and $p=0.515$.  Inset: same results plotted with different range to show the initial ($t<10\,t_s$) behaviour. Results analyzed using the density threshold condition $\epsilon=0.1$.}
   \label{fig:qdep}
\end{figure}
 
\subsubsection{Dependence on quadratic Zeeman energy}\label{Secq}
Our results thus far have been presented for the case of $q=-0.3q_0$. The quadratic Zeeman energy shifts the $m=\pm1$ Zeeman sublevels relative to the $m=0$ sublevel. For small negative values of the quadratic Zeeman energy ($-q_0<q<0$) interactions are able to drive the spin-mixing of atoms back into the $m=0$ level in the initial unstable dynamics. We observed this as a dominant source of the variation in $p_c$ with time in Fig.~\ref{fig:td_domains}.

For larger negative values ($q\le -q_0$) spin-mixing is energetically unfavourable. Indeed, an analysis of the quasi-particle excitations on the initial state (for $\varphi=0$) is shown in Figs.~\ref{fig:qdep}(a) and (b) for the cases $q=-0.3q_0$ and $-q_0$, respectively. There are three excitation branches with dispersion relations $\epsilon_\mu(k)$, which we label as $\mu=\{0,1,2\}$ (see \cite{Symes2014}), with analytically known properties for the $\varphi=0$ case \cite{Kawaguchi2012R}.  
For $q=-0.3q_0$ the magnon branches $\epsilon_1$ and $\epsilon_2$  are both dynamically unstable\footnote{The phonon branch, $\epsilon_0(k)$, associated with density functions is stable as $g_n>0$.}, i.e.~with imaginary energies [$\mathrm{Im}\{\varepsilon_\mu(k)\}\ne0$] for a range of excitation wavevectors $k$. Such dynamically unstable modes will exponentially grow with time, and give rise to magnetized domains. The $\epsilon_2$ branch consists of ``transverse magnons'', which have amplitude in $m=0$ sublevel. When these excitations grow on top of the condensate (noting the condensate is mainly in the $m=\pm1$ sublevels) it leads to the formation of transverse magnetization $\mathbf{F}_\perp=(F_x,F_y)$. Significantly, the growth of these modes corresponds to spin-mixing of atomic population into the $m=0$ sublevel. The $\epsilon_1$ excitation branch consists of ``axial magnons'', which have amplitude in the $m=\pm1$ sublevels. As these modes grow they cause the population in the $m=\pm1$ to spatially separate into positive and negative domains, i.e.~these modes directly initiate the immiscibility dynamics leading to the formation of the EA $F_z$ domains. For $q=-q_0$ [ Fig.~\ref{fig:qdep}(b)] only the axial magnons are dynamically unstable, and in this case there will be no spin-mixing.

An example contrasting the quenches to $q=-0.3q_0$ and $-q_0$ is shown in Fig.~\ref{fig:qdep}(d). We see that  the $N_+$ population for the $-q_0$ quench is constant (c.f.~$-0.3q_0$, where spin-mixing causes $N_+$ to dip by about $20\%$), and as a result the domain area  $A_+$ is  seen to saturate faster, and vary more slowly in time, than the $q=-0.3q_0$ case.
As a result the percolation analysis (in terms of $p$) shows that   $p_c$ for the $q=-q_0$ case has a much weaker time dependence than for the $q=-0.3q_0$ case. 
Deeper $q$ quenches will not lead to any additional changes in the dynamics, as the $m=0$ level [already absent from our initial condition (\ref{EqInitStateQ2})], will remain frozen out from the dynamics. Indeed, in this regime the system behaves as a  two-component system.

\begin{figure}[htbp] 
   \centering 
    \includegraphics[width=5.75in]{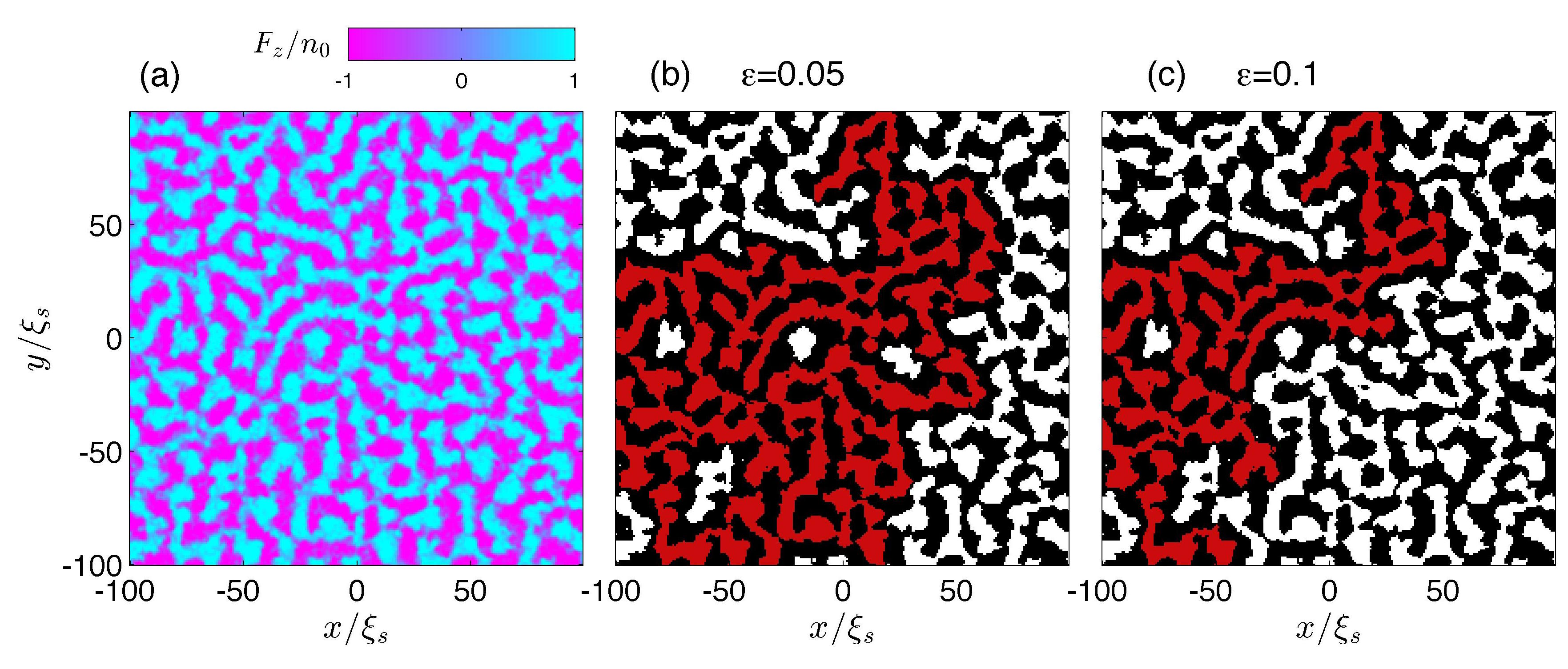} 
    \includegraphics[width=5.75in]{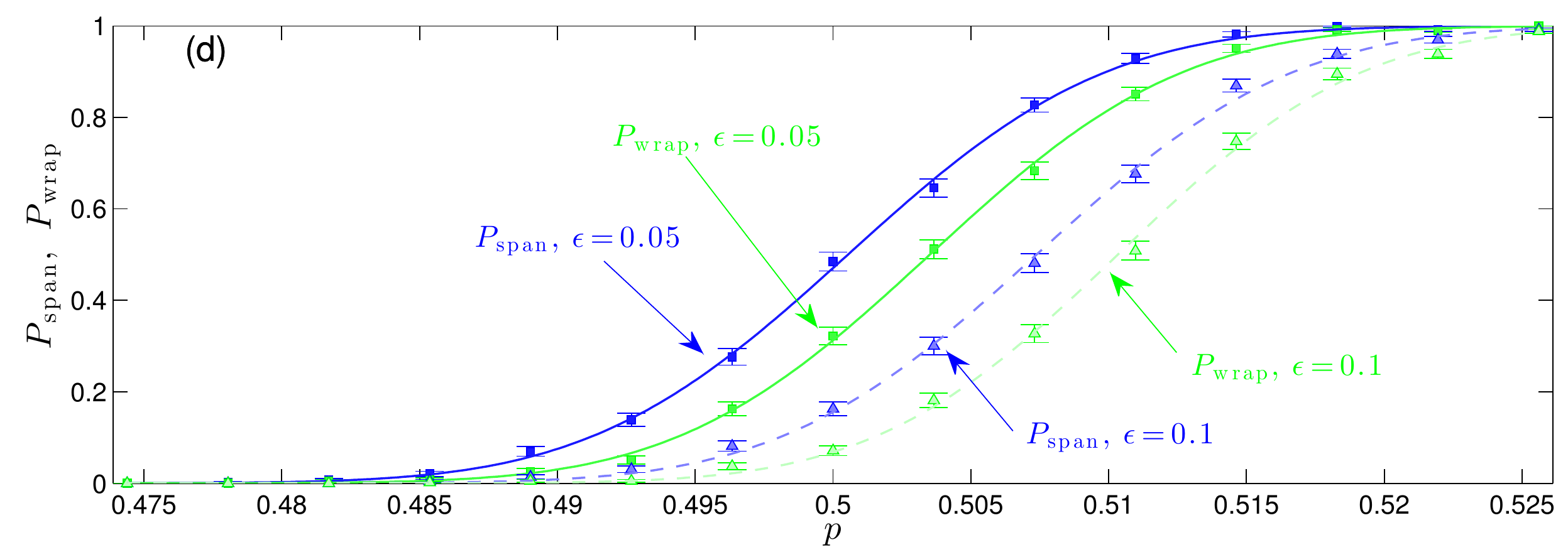} 
   \caption{Sensitivity of domain analysis to the density threshold condition $\epsilon$. (a)  The $z$-spin density for $p=0.5$ at $t=50\,t_s$ after the quench. 
     The binary images obtained from (a)  using (b) $\epsilon=0.05$ and (c) $\epsilon=0.1$. (d) Probability that at least one domain is spanning (blue) or wrapping (green) at $t=50\,t_s$ for a system of size $L=200\,\xi_s$ with $N_L=256$ points in each direction for $\epsilon=0.05$ (squares) and  $\epsilon=0.1$ (triangles). 
   Lines are the fits to the numerical results using Eq.~(\ref{Eq:pfit}).  
     Simulations for $n_0=10^4/\xi_s^2$, $q=-0.3\,q_0$, $g_s=-\frac{1}{3}g_n$, $L=200\,\xi_s$, $N_L=256$,  and analyzed using $N_{\mathrm{traj}}=590$}
   \label{fig:edep}
\end{figure}

\subsubsection{Sensitivity to the density threshold condition}\label{Sece}
Our analysis thus far has been based on defining positive domains via the density threshold condition (\ref{eq:sigmadefn}) with $\epsilon=0.1$. We now consider the effect of changing $\epsilon$. In Fig.~\ref{fig:edep}(a) we show the $F_z$ density of a domain produced from a simulation with $p=0.5$. Figures~\ref{fig:edep}(b) and (c) show the binary image  and largest domain constructed from that result using  $\epsilon=0.05$ and $\epsilon=0.1$, respectively. Most noticeably the extent of the largest domain reduces with increasing $\epsilon$. This is because often the domains are connected by tenuous regions with $|F_z|\ll n_0$, which are sensitive to the threshold condition.

 In Fig.~\ref{fig:edep}(d) we present results for the spanning and wrapping probabilities for these two values of $\epsilon$. The probability curves shift to the right for increasing $\epsilon$, with the values of 
$p_c^{\mathrm{eff}}(L)$ [from fitting the data to (\ref{Eq:pfit})] for the spanning (wrapping) results being $\{0.498,0.505\}$  
 ($\{0.502,0.508\}$)  
 for $\epsilon=\{0.05,0.1\}$, i.e.~about a $1\%$ change.

\section{Outlook and Conclusions}\label{SecConclusion}

\begin{figure}[htbp] 
   \centering 
    \includegraphics[width=5.5in]{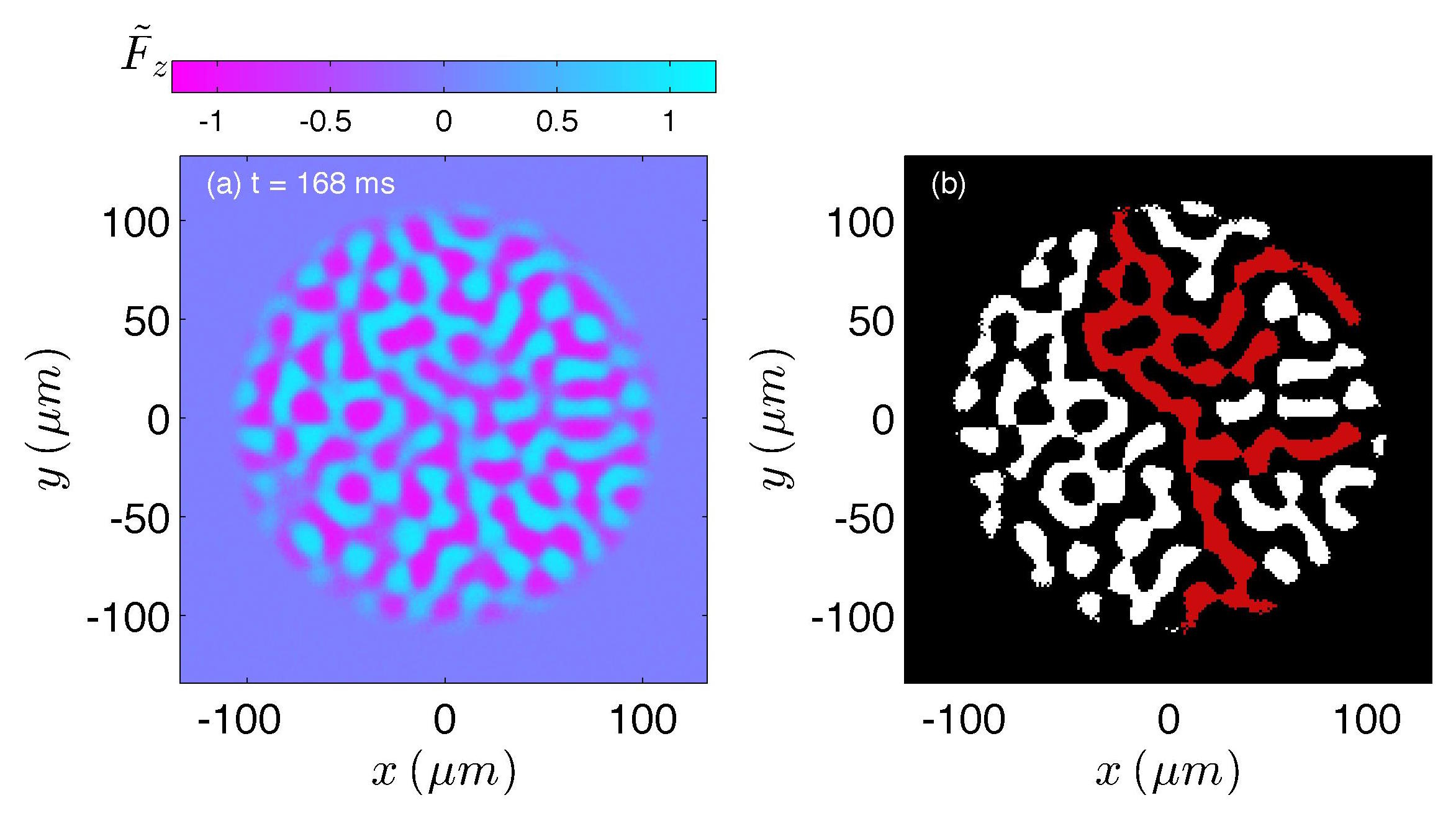} 
   \caption{Example of domains in a trapped system using $^{87}$Rb parameters. (a) Normalized magnetization column density $\tilde{F}_z\equiv  \int dz F_z(\mathbf{x})\,/n_{\mathrm{2D}}^{\mathrm{peak}}$, where $n_{\mathrm{2D}}^{\mathrm{peak}}$ is the peak (areal) column density of the initial condensate. (b) The binary image constructed using $\tilde{F}_z>\epsilon$, with $\epsilon=0.1$ as the threshold condition. Positive domains are shaded white, except for the largest domain which is shaded red. Simulation for $p=0.5$ case, with $q=-0.5q_0$. Condensate of $6\times10^6$ atoms in flat trap: $V(\mathbf{x})=V_\rho(\sqrt{x^2+y^2})+\frac{1}{2}M\omega_z^2z^2$,  where $V_\rho(\rho)=\frac{1}{2}V_0\{\tanh[(\rho-R)/w_\rho]+1\}$, with $\omega_z/2\pi=10^3\,$s$^{-1}$, $V_0/h=4\times10^3\,$s$^{-1}$, $R=114\,\mu $m and $w_\rho=7.6\,\mu $m. Using the peak initial condensate density $n^{\mathrm{peak}}$ we have $q_0/h=14.2\,$Hz, $\xi_s=2.9\,\mu $m and $t_s=19.9\,$ms. Initial preparation: a ground state condensate for potential $V(\mathbf{x})$ is produced with all atoms in $m=0$ state. This state has  $n_{\mathrm{2D}}^{\mathrm{peak}}=1.75\times10^{14}\,$m$^{-2}$  and $n^{\mathrm{peak}}=1.97\times10^{20}\,$m$^{-3}$. Vacuum Wigner noise is added to planar momentum modes $\mathbf{k}_{\rho}=(k_x,k_y)$ on the condensate, restricted to $ {\hbar^2k_{\rho}^2}/{2M}< \hbar\omega_z$. This noise is projected onto the spatial region inside the trap, i.e.~$\sqrt{x^2+y^2}<R$. The spin rotations described in Sec.~\ref{SecInitState} are applied to this condensate with noise to produce the initial condition for the simulation. } 
   \label{fig:trapdomains}
\end{figure}

  {
In this paper we have investigated the percolation of EA magnetic domains forming in spinor condensate. This could be investigated using a quench of  the spin-dependent interaction, however we have instead proposed a scheme using a quadratic Zeeman quench and a genrealized spin rotation applied to a ferromagnetic spin-1 condensate, which will be feasible to implement in current experiments. By varying the conserved magnetization $M_z$ of the initial state, and hence the proportion of positive and negative EA domains, this system is able to explore the percolation transition.} 
Using an ensemble of simulations of a quasi-2D system we have quantified the probability of percolation occurring as $M_z$ is varied, and for systems of various sizes. From these results we use finite-size scaling to extract the correlation length critical exponent, obtaining a value consistent with standard 2D percolation. We also use finite size scaling to extrapolate to the infinite system percolation  threshold.   

We have explored various aspects of the system dynamics, including the role of spin-mixing in the early time evolution of the percolating clusters, showing that the percolation threshold is time dependent. 
We showed that this effect can be mitigated by directly measuring the positive domain area to define an occupation probability $p'$, or by quenching to deeper values of the quadratic Zeeman energy, where spin-mixing is suppressed.
At late times ($t>10^2\,t_s$), not considered here, the system will begin to phase order, and the typical size of domains will grow as $l_d(t)\sim t^{2/3}$ \cite{Kudo2013a,Williamson2016b}. In this regime the system should exhibit the phenomenon of ``phase-ordering percolation'' \cite{Takeuchi2015a}, whereby the effective size of the system diminishes in time as $L/l_d(t)$. This would be an interesting direction for future investigation.

Our results indicate that it may be viable to study percolation in experiments. One important reason is that the domains form in the early time dynamics [i.e.~$t\sim O(10\,t_s)$], which is a time scale easily accessible to experiments with $^{87}$Rb (ferromagnetic) spinor condensates, which have a small value of $|g_s|$ and hence $t_s$ is large.   {Also it is feasible to manipulate the quadratic Zeeman energy on such time-scales with negligible heating (e.g.~see \cite{Luo2017a})}. A full study of the experimental system is outside the scope of this paper, nevertheless it is useful to explore the feasibility of observing EA domains in a  realistic scenario. We indicate a result from such a calculation in Fig.~\ref{fig:trapdomains}, showing the domains formed and a domain that vertically spans the system. This result is constructed from the column density of a $^{87}$Rb condensate in a flat-bottomed trap. Importantly this result shows that it is possible to get a reasonable number of domains forming, such that percolation properties will be nontrivial.  
It would also be interesting to consider a highly oblate harmonic trap. However, as the density decreases as we move towards the edge of such a trap the spin healing length, and hence the typical domain size, will also increase. Another area for future exploration is the role of finite temperature, whereby the initial state will have thermally occupied excitations. This will cause the domains to form more rapidly, but could also influence the nature of the domains that form.

\ack\addcontentsline{toc}{section}{Acknowledgments}
The authors acknowledge support from the Marsden Fund of the Royal Society of New Zealand.   PBB thanks Yongyong~Cai for his support of this research, and would like to acknowledge useful discussions with Russell~Anderson, Lincoln~Turner and Niels~Kj{\ae}rgaard about experimental methods for preparing $^{87}$Rb into the initial state.
 
\footnotesize
\vspace*{0.25cm} 

\addcontentsline{toc}{section}{References}





 \providecommand{\newblock}{}

\end{document}